\documentclass[12pt,leqno]{article}
\title{Logotropic distributions }

\usepackage{amsthm,amsmath,amssymb,a4wide,psfig,epsf}
\def\mb#1{\setbox0=\hbox{$#1$}\kern-.025em\copy0\kern-\wd0
\kern-0.05em\copy0\kern-\wd0\kern-.025em\raise.0233em\box0}

\begin{document}

\author{Pierre-Henri Chavanis and Cl\'ement Sire}
\maketitle
\begin{center}
Laboratoire de Physique Th\'eorique (UMR 5152 du CNRS), Universit\'e Paul Sabatier,\\
118, route de Narbonne, 31062 Toulouse Cedex, France\\ E-mail:
{\it chavanis{@}irsamc.ups-tlse.fr \&
clement.sire{@}irsamc.ups-tlse.fr }

%\date{}
\vspace{0.5cm}
\end{center}

\begin{abstract}

In all spatial dimensions $d$, we study the static and dynamical
properties of a generalized Smoluchowski equation which describes the
evolution of a gas obeying a logotropic equation of state,
$p=A\ln\rho$. A logotrope can be viewed as a limiting form of
polytrope ($p=K\rho^{\gamma}$, $\gamma=1+1/n$), with index $\gamma=0$
or $n=-1$. In the language of generalized thermodynamics, it
corresponds to a Tsallis distribution with index $q=0$. We solve the
dynamical logotropic Smoluchowski equation in the presence of a fixed
external force deriving from a quadratic potential, and for a gas of
particles subjected to their mutual gravitational force. In the latter
case, the collapse dynamics is studied for any negative index $n$, and
the density scaling function is found to decay as $r^{-\alpha}$, with
$\alpha=\frac{2n}{n-1}$ for $n<-\frac{d}{2}$, and
$\alpha=\frac{2d}{d+2}$ for $-\frac{d}{2}\leq n<0$. 

\vskip0.5cm


\end{abstract}

\section{Introduction}
\label{intro}

Recently, several researchers have studied generalized forms of
Fokker-Planck equations \cite{collpap,frankbook}. These equations arise when
the coefficients of diffusion, mobility and friction in the usual
Fokker-Planck equations (Kramers, Smoluchowski,...) explicitly
depend on the concentration of particles. This can take into
account phenomenologically ``hidden constraints'' that are not
directly accessible to the observer \cite{blks}.  These nonlinear
Fokker-Planck equations are associated with generalized free
energy functionals and non-standard distributions. For example,
Tsallis $q$-distributions constitute an important class of
non-Boltzmannian distributions \cite{tsallis}.  Plastino \&
Plastino \cite{pp} have shown that they could be obtained as
stationary solutions of a nonlinear Fokker-Planck equation, with
potential applications to the context of porous media. Recently,
Kaniadakis \cite{kaniadakis}, Frank \cite{frank} and Chavanis
\cite{gen} have shown that it was possible to construct more
general Fokker-Planck equations leading to other forms of
non-standard distributions. For example, Chavanis
\cite{gen,next03,banach} has introduced a generalized Smoluchowski
equation involving an arbitrary barotropic equation of state
$p=p(\rho)$. This can be derived from a generalized Kramers
equation in a strong friction limit \cite{gen,lemou}.  The usual
Smoluchowski equation leading to the Boltzmann distribution
corresponds to an isothermal equation of state $p=\rho T$ while
the Tsallis distributions are associated with a polytropic
equation of state $p=K\rho^{\gamma}$ where $\gamma=1+\frac{1}{n}$
plays the role of the Tsallis $q$-parameter.

In this paper, we consider the case of a logotropic equation of state
$p=A\ln\rho$ where $A$ is a positive constant. This equation of state
sometimes appears in astrophysics \cite{log}, but we consider it here
at a more general level. In Sec. \ref{sec_nmfp}, we show that a
logotrope can be viewed as a limiting form of a polytrope with
$\gamma=0$ (or as a Tsallis distribution with $q=0$). In
Sec. \ref{sec_qp}, we solve the logotropic Smoluchowski equation
with a fixed quadratic potential.  This is a particular case of the
general analytical solution found by Plastino \& Plastino \cite{pp} and Tsallis
\& Buckman \cite{tb} for the nonlinear Fokker-Planck equation
(associated with polytropes).  In Sec. \ref{sec_lsp}, we consider a coupling with
the gravitational potential \cite{ts,lang} and study the logotropic
Smoluchowski-Poisson (LSP) system. This is a particular case of the
nonlinear (polytropic) Smoluchowski-Poisson system studied in
\cite{lang}, but with a negative index $n=-1$ . We characterize the
stationary solutions of the LSP system, determine their stability and
find self-similar collapse solutions when no equilibrium state exists
(in a space of dimension $d>2$). We find that the collapse of
polytropes with negative index is peculiar: (i) for $n<-d/2$, the
scaling exponent is $\alpha=2n/(n-1)$ like for positive index (ii) for
$-d/2<n<0$, the scaling exponent is $\alpha=2d/(d+2)$
corresponding to the collapse of a cold (pressureless) system. In
particular, the scaling exponent of logotropes is $\alpha=2d/(d+2)$
($d>2$).

\section{Nonlinear mean-field Fokker-Planck equations} \label{sec_nmfp}

\subsection{The generalized Smoluchowski equation} \label{sec_gs}

In a recent series of papers
\cite{lang,crs,sc,iso,hmf,multi,bose,virial,tr}, we have studied a
general class of nonlinear mean-field Fokker-Planck equations of the
form
\begin{equation}
\label{gs1} \frac{\partial\rho}{\partial t}=\nabla\cdot \left
\lbrack\frac{1}{\xi}\left (\nabla p+\rho\nabla\Phi \right )\right
\rbrack,
\end{equation}
\begin{equation}
\label{gs2}\Phi({\bf r},t)=\int u(|{\bf r}-{\bf r}'|)\rho({\bf
r}',t)d{\bf r}',
\end{equation}
introduced in Chavanis \cite{gen,next03,banach,hb}.  This type of equations arises in
different domains of physics, chemistry, astrophysics and biology
\cite{cras}. The drift-diffusion equation (\ref{gs1}) can be
viewed as a generalized Smoluchowski equation where $p=p(\rho)$ is
a barotropic equation of state and $\Phi({\bf r},t)$ is a
potential. This potential can be imposed by an external medium (in
which case $\Phi=\Phi_{ext}({\bf r})$ is assumed given) or
generated by the particles themselves through the mean-field
equation (\ref{gs2}) where $u(|{\bf r}-{\bf r}'|)$ is a binary
potential of interaction. The generalized Smoluchowski equation
(\ref{gs1}) can be derived from a generalized Kramers equation in
a strong friction limit $\xi\rightarrow +\infty$ by performing a
Chapman-Enskog expansion in terms of the small parameter $1/\xi$
\cite{lemou} or by using a method of moments \cite{banach}. It can
also be directly obtained from the damped barotropic Euler
equations by neglecting the inertial term in the equation for the
velocity \cite{gen}. Thus, the generalized Smoluchowski equation
can be interpreted as an overdamped limit of kinetic or
hydrodynamic equations in which a friction force (real or
effective) is present \cite{virial}.

The generalized mean-field Smoluchowski system
(\ref{gs1})-(\ref{gs2}) decreases the Lyapunov functional
\begin{equation}
\label{gs3}
F[\rho]=\int\rho\int^{\rho}\frac{p(\rho')}{\rho^{'2}}d\rho' \,
d{\bf r}+\frac{1}{2}\int\rho\Phi \, d{\bf r},
\end{equation}
which can be interpreted as a generalized free energy \cite{gen}. Indeed, $\dot F=-\int {1\over\xi\rho}(\nabla p+\rho\nabla \Phi)^{2}d{\bf r}\le 0$.
The stationary solutions of Eq.~(\ref{gs1}) satisfy the condition
\begin{equation}
\label{gs4}
\nabla p+\rho\nabla\Phi={\bf 0},
\end{equation}
which can be viewed as a condition of hydrostatic equilibrium.  Using
Eq.~(\ref{gs4}) and the equation of state $p=p(\rho)$, we find that
the equilibrium density field is of the form $\rho=\rho\lbrack
\Phi({\bf r})\rbrack$. When substituted in Eq. (\ref{gs2}), this
yields an integro-differential equation for the potential $\Phi({\bf
r})$. The steady states of Eqs.  (\ref{gs1})-(\ref{gs2}) are extrema
of the free energy (\ref{gs3}) at fixed mass $M=\int \rho \, d{\bf
r}$. They satisfy the first order variations $\delta F-\alpha\delta
M=0$ which directly return Eq.~(\ref{gs4}). Furthermore, the steady
states are linearly dynamically stable if, and only if, they are {\it
minima} of the free energy (\ref{gs3}) at fixed mass $M=\int \rho \,
d{\bf r}$ \cite{gen}.

When $\Phi$ is interpreted as the gravitational potential determined
by the Poisson equation, the {\it stationary} solutions of the
generalized Smoluchowski-Poisson system (GSP) are the same as the
stationary solutions of the Euler-Poisson system
\begin{equation}
\label{gs5} \frac{\partial\rho}{\partial t}+\nabla (\rho {\bf u})=0,
\end{equation}
\begin{equation}
\label{gs6} \frac{\partial {\bf u}}{\partial t}+({\bf u}\cdot \nabla){\bf u}=
-\frac{1}{\rho}\nabla p+\nabla \Phi,
\end{equation}
\begin{equation}
\label{gs7}\Delta\Phi=S_{d}G\rho,
\end{equation}
describing barotropic stars \cite{bt}. However, the {\it dynamics} of these two
systems is different as the Euler-Poisson system describes a fluid
without friction $\xi=0$ while the generalized Smoluchowski-Poisson
system describes a Brownian gas in an overdamped limit $\xi\rightarrow
+\infty$ \cite{virial}. With these analogies and differences in mind,
we can use many results obtained in astrophysics \cite{bt} when
studying nonlinear mean-field Fokker-Planck equations of the form
(\ref{gs1})-(\ref{gs2}). We think therefore that it is enlightening to
use a similar vocabulary and similar notations.

We note, finally, that drift-diffusion equations of the form
(\ref{gs1})-(\ref{gs2}) also arise in biology, in the context of
chemotaxis \cite{iso,jtype}. They can be written in the form
\begin{equation}
\label{gs8} \frac{\partial\rho}{\partial t}=\nabla\cdot \left
\lbrack\chi\left (\nabla p-\rho\nabla c\right )\right
\rbrack,
\end{equation}
\begin{equation}
\label{gs9}\frac{\partial c}{\partial t}=-k(c)c+f(c)\rho+D_{c}\Delta c.
\end{equation}
They describe the collective motion of cells (usually bacteria or
amoebae) that diffuse and that are attracted by a chemical substance
that they emit. Here, $\rho({\bf r},t)$ denotes the cell density and
$c({\bf r},t)$ is the concentration of chemo-attractant which induces
the drift force. This model has been introduced by Keller \& Segel
\cite{ks}. In its most studied version \cite{jaeger},
$p(\rho)=(D_{*}/\chi)\rho$ is a linear function of the density which
is similar to an isothermal equation of state. However, the general
Keller-Segel model \cite{ks} allows the possibility that the diffusion
be anomalous so that the function $p=p(\rho)$ can be nonlinear. Note
also that in the limit of large diffusivity of the chemical
$D_{c}\rightarrow +\infty$ (and for $k(c)=k$ and $f(c)=f$), the time
derivative $\partial c/\partial t$ can be neglected \cite{jaeger} so
that Eq. (\ref{gs9}) reduces to a Poisson equation (like in gravity)
with a screening term $-kc$.  Therefore, the general study of
equations of the form (\ref{gs1})-(\ref{gs2}) connects several active
areas of research in theoretical physics: nonlinear Fokker-Planck
equations, self-gravitating systems, chemotaxis and non-extensive
thermostatistics. Here, we shall focus on the case where the equation
of state is that of a logotrope $p=A\ln\rho$ and we shall make the
connection with the previously studied polytropes $p=K\rho^{\gamma}$.

\subsection{The polytropic Smoluchowski equation} \label{sec_polys}

For the  polytropic equation of state $p=K\rho^{\gamma}$ (with
$\gamma=1+1/n$), we obtain the polytropic Smoluchowski  equation
\begin{equation}
\label{polys1} \frac{\partial\rho}{\partial t}=\nabla\cdot \left
\lbrack\frac{1}{\xi} \left (K\nabla \rho^{\gamma}+\rho\nabla\Phi
\right )\right \rbrack.
\end{equation}
The free energy (\ref{gs3}) corresponding to a polytropic equation of
state can be conveniently written
\begin{equation}
\label{polys2} F[\rho]=\frac{K}{\gamma -1}\int
(\rho^{\gamma}-\rho) \, d{\bf r} +\frac{1}{2}\int\rho\Phi \, d{\bf
r},
\end{equation}
and the stationary solutions of Eq.~(\ref{polys1}) are given by
\begin{equation}
\label{polys3} \rho=\left \lbrack
\lambda-\frac{\gamma-1}{K\gamma}\Phi \right
\rbrack^{\frac{1}{\gamma-1}}.
\end{equation}
This polytropic distribution reproduces the statistics introduced
by Tsallis in his generalized thermodynamics \cite{tsallis}. When
$\Phi$ is a fixed external potential, the polytropic Smoluchowski
equation (\ref{polys1}) is equivalent to the nonlinear
Fokker-Planck equation introduced by Plastino \& Plastino
\cite{pp}. When $\Phi=0$, it returns the equation of porous media.
We note that the free energy (\ref{polys2}) can be written
$F=E-T_{\it eff} S$ with $E=\frac{1}{2}\int \rho\Phi \, d{\bf r}$,
$T_{\it eff}=K$ and $S=-\frac{1}{\gamma -1}\int
(\rho^{\gamma}-\rho) d{\bf r}$. In the language of generalized
thermodynamics, this can be viewed as a Tsallis free energy with
index $\gamma$ where $K$ plays the role of a generalized
temperature \cite{gen}.  For $\gamma=1$, i.e. $n=+\infty$, we
recover the isothermal equation of state $p=\rho T$ and the
Boltzmann free energy $F=\frac{1}{2}\int \rho\Phi \, d{\bf
r}+T\int \rho\ln\rho\, d{\bf r}$. As we shall see, the case
$\gamma=0$, i.e. $n=-1$, is also special and corresponds to what
have been called logotropes in astrophysics \cite{log}.

\subsection{The logotropic Smoluchowski equation} \label{sec_logs}

Let us consider the logotropic equation of state
\begin{equation}
\label{logs1}
p=A\ln \rho,
\end{equation}
where $A$ is a constant.  This equation of state has been
introduced in astrophysics to account for certain properties of
molecular clouds that could not be understood in terms of
isothermal distributions \cite{log}. However, this equation of
state can have application in more general situations, beyond the
realm of astrophysics, so we shall consider it here on a general
footing. Inserting Eq.~(\ref{logs1}) in Eq.~(\ref{gs1}), we get
the logotropic Smoluchowski equation
\begin{equation}
\label{logs2} \frac{\partial\rho}{\partial t}=\nabla\cdot \left
\lbrack\frac{1}{\xi}\left (A\nabla \ln\rho+\rho\nabla\Phi \right
)\right \rbrack.
\end{equation}
The free energy (\ref{gs3}) of a logotrope takes the form
\begin{equation}
\label{logs3} F[\rho]=-A\int \ln\rho \, d{\bf
r}+\frac{1}{2}\int\rho\Phi \, d{\bf r}.
\end{equation}
It can be written $F=E-T_{\it eff}S$ with $T_{\it eff}=A$ and
\begin{equation}
\label{logs4} S[\rho]=\int \ln\rho \, d{\bf r}.
\end{equation}
In a previous paper \cite{super}, this functional has been called
the log-entropy. The stationary states of the logotropic
Smoluchowski equation are given by
\begin{equation}
\label{logs5}
\rho=\frac{1}{\lambda+\frac{1}{A}\Phi}.
\end{equation}
Note that when the potential is quadratic, i.e.  $\Phi=r^{2}/2$, the
equilibrium distribution is the Lorentzian as noted in \cite{super}. Thus, the
log-entropy (\ref{logs4}) can be seen as the functional associated with the
Lorentzian distribution.

\subsection{The polytropic index $\gamma=0$} \label{sec_z}

Returning to the polytropic Smoluchowski equation (\ref{polys1}), we
note that the index $\gamma=0$ is special because the polytropic
Smoluchowski equation (\ref{polys1}) and the Tsallis free energy
(\ref{polys2}) become trivial for this index except if $K\rightarrow
+\infty$ \cite{super}. This type of indetermination is often
characteristic of a logarithmic behavior as can be seen with a simple
change of variables. If we set $K=A/\gamma$, the polytropic
Smoluchowski equation (\ref{polys1}) can be rewritten
\begin{equation}
\label{z1} \frac{\partial\rho}{\partial t}=\nabla\cdot  \left
\lbrack\frac{1}{\xi}\left (A \rho^{\gamma-1}\nabla
\rho+\rho\nabla\Phi \right )\right \rbrack.
\end{equation}
Under this form, this equation describes polytropes for
$\gamma\neq 0$ and logotropes for $\gamma=0$. Therefore, a
logotrope can be viewed as a polytrope with index $\gamma=0$, i.e.
$n=-1$.  The stationary solution of equation (\ref{z1}) is given
by
\begin{equation}
\label{z2} \rho=\left \lbrack \lambda-\frac{\gamma-1}{A}\Phi\right
\rbrack^{\frac{1}{\gamma-1}},
\end{equation}
and this distribution passes to the limit for $\gamma\rightarrow
0$. Indeed, for $\gamma\neq 0$ we can put Eq.~(\ref{z2}) in the
form of Eq.~(\ref{polys3}) and for $\gamma=0$, we recover
Eq.~(\ref{logs5}). The free energy associated with Eq.~(\ref{z1})
can be written
\begin{equation}
\label{z3} F[\rho]=\frac{A}{\gamma(\gamma -1)}\int  \left
(\rho^{\gamma}-\rho\right ) \, d{\bf r}+\frac{1}{2}\int\rho\Phi \,
d{\bf r}.
\end{equation}
For $\gamma\neq 0$, we can put Eq.~(\ref{z3}) in the form of
Eq.~(\ref{polys2}) and for $\gamma=0$, we recover
Eq.~(\ref{logs3}) up to additive terms that do not depend on the
density. Since only the variations of $F$ matter, these terms
(that can be infinite!) do not play any role \cite{super}. In the
same spirit, the general polytropic equation of state can be
written $p=\frac{A}{\gamma}\rho^{\gamma}$ with the convention that
$\lim_{\gamma\rightarrow 0}\frac{\rho^{\gamma}}{\gamma}=\ln
\rho+{\rm Cst}$. {\it Therefore, in the language of generalized
thermodynamics, logotropes correspond to Tsallis statistics with
index $\gamma=0$.}

\subsection{Other formulations} \label{sec_form}

We can write the generalized Smoluchowski equation in the form
\begin{equation}
\label{form1} \frac{\partial \rho}{\partial t}=\nabla\cdot
\left\lbrack D\left  (\rho C''(\rho)\nabla\rho+\beta
\rho\nabla\Phi\right )\right \rbrack,
\end{equation}
where $C(\rho)$ is a convex function, $\beta$ is a constant which
can be interpreted as a generalized inverse temperature and $D$ is a
positive constant which can be interpreted as a generalized
diffusion coefficient. The mobility $\mu=1/\xi$ satisfies a form of
Einstein relation $\mu=D\beta$. From the convex function $C$ we
define a generalized entropic functional
\begin{equation}
\label{form2} S[\rho]=-\int C(\rho)\, d{\bf r},
\end{equation}
and a generalized free energy $J=S-\beta E$. This functional
increases monotonically with time ($\dot J=\int {D\over\rho}(\rho C''(\rho)\nabla\rho+\beta\rho\nabla\Phi)^{2}d{\bf r}\ge 0$). This can be
viewed as a form of $H$-theorem in a canonical description where
the inverse temperature $\beta$ is fixed \cite{gen}. The steady
solutions of Eq.~(\ref{form1}) are determined by the
integro-differential equation
\begin{equation}
\label{form3} C'(\rho)=-\beta\Phi-\alpha,
\end{equation}
where $\Phi$ is related to $\rho$ by Eq.~(\ref{gs2}) in the
general case. They are critical points of $J$ at fixed mass $M$;
indeed, the first order variations satisfy $\delta J-\alpha\delta
M=0$ which returns Eq.~(\ref{form3}). Furthermore, the linearly
dynamically stable stationary solutions of Eq.~(\ref{form1}) are
maxima of $J$ at fixed mass $M$.

For the Tsallis entropic functional written in the form
\begin{equation}
\label{form4} S[\rho]=-\frac{1}{\gamma (1-\gamma)}\int
(\rho^{\gamma}-\rho)\, d{\bf r},
\end{equation}
the generalized Smoluchowski equation (\ref{form1}) reads
\begin{equation}
\label{form5} \frac{\partial \rho}{\partial t}=\nabla\cdot
\left\lbrack  D\left (\rho^{\gamma-1}\nabla\rho+\beta
\rho\nabla\Phi\right )\right \rbrack.
\end{equation}
Comparing with Eq.~(\ref{z1}), we find that $\beta=1/A$. This is
consistent with the previous formalism where $A=T_{eff}$ is
interpreted as an effective temperature. The stationary solution
of Eq.~(\ref{form5}) can be written
\begin{equation}
\label{form6}
\rho=\left \lbrack \lambda-\beta(\gamma-1)\Phi\right\rbrack^{\frac{1}{\gamma-1}}.
\end{equation}
We note that the parameter $\beta$ has not the dimension of an inverse
temperature, although this is the quantity which naturally enters in
the free energy functional $J$ and in the variational principle
$\delta S-\beta\delta E-\alpha\delta M=0$ determining the stationary
solution (\ref{form6}). As discussed in \cite{intT}, there are
different notions of temperature in the case of polytropic
distributions. We can write the distribution (\ref{form6}) in the alternative
form
\begin{equation}
\label{form7}
\rho=Q\left \lbrack 1-b(\gamma-1)\Phi\right\rbrack^{\frac{1}{\gamma-1}},
\end{equation}
where $Q=\lambda^{1\over \gamma-1}$ and $b=\beta/\lambda$ now has the
dimension of an inverse  temperature.

\subsection{The logotropic Kramers and Landau equations} \label{sec_kl}

The generalized Fokker-Planck equations (\ref{gs1}) and
(\ref{form1}) are written in position space. We can also introduce
generalized Fokker-Planck equations in velocity space. When the
friction term is linear in ${\bf v}$, they are referred to as
generalized Kramers equations.  Some exemples are given in
\cite{gen} for different forms of entropic functionals $S=-\int
C(f)\,d{\bf r}d{\bf v}$. For the log-entropy $C(f)=-\ln f$
considered in this paper, we obtain the logotropic Kramers
equation
\begin{equation}
\label{kl1} \frac{\partial f}{\partial t}=\frac{\partial}{\partial
{\bf v}} \cdot \left \lbrack D\left ( \frac{\partial \ln
f}{\partial {\bf v}}+\beta f {\bf v}\right )\right\rbrack.
\end{equation}
The  generalized Kramers equation is associated with a {\it
canonical description} where the temperature $\beta$ is fixed and
the free energy $J=S-\beta E$ increases \cite{gen,hb}. Alternatively, we can
consider the generalized Landau equation that is associated with a
{\it microcanonical description}  where the energy $E=\int f
{v^2\over 2}d{\bf r}d{\bf v}$ is conserved and the generalized entropy
$S$ increases \cite{blks,hb}. For the log-entropy $C(f)=-\ln f$, we obtain the
logotropic Landau equation
\begin{equation}
\label{kl2} \frac{\partial f}{\partial
t}=A\frac{\partial}{\partial v^{\nu}}  \int
\frac{u^{2}\delta^{\mu\nu}-u^{\mu}u^{\nu}}{u^{3}}\left\lbrack
f'\frac{\partial \ln f}{\partial v^{\nu}}-f\frac{\partial \ln
f'}{\partial v^{'\nu}}\right\rbrack\, d{\bf v}',
\end{equation}
where $f=f({\bf v},t)$ and $f'=f({\bf v}',t)$. The steady solution
of the logotropic Kramers and Landau equations is the Lorentzian
\begin{equation}
\label{kl3}
f_{e}=\frac{1}{\lambda+\beta\frac{v^{2}}{2}}.
\end{equation}
Note that if the velocity is not bounded, this distribution is not
normalizable in $d\ge 2$.

\section{The logotropic Smoluchowski equation with a fixed quadratic potential} \label{sec_qp}

We shall here provide an analytical time-dependant solution of the
logotropic Smoluchowski equation in $d=1$ for a fixed quadratic
potential $\Phi=x^{2}/2+\lambda x$. This solution is a
straightforward extension of the solution found by Plastino \&
Plastino \cite{pp} and Tsallis \& Bukman \cite{tb} for the
polytropic Smoluchowski equation with $\gamma\neq 0$.  We cannot
directly make $\gamma=0$ in their solution for the reason
discussed in Sec. \ref{sec_z}. However, if we write the polytropic
Smoluchowski equation in the form
\begin{equation}
\label{qp1} \frac{\partial \rho}{\partial
t}=\frac{\partial}{\partial x} \left
\lbrack\rho^{\gamma-1}\frac{\partial \rho}{\partial x}+\beta\rho
x+\lambda \rho\right\rbrack,
\end{equation}
we can follow exactly the reasoning of \cite{pp,tb} and finally
let $\gamma\rightarrow 0$ since this limit is now well-behaved.
Note that Eq.~(\ref{qp1}) with $\lambda=0$ coincides with the
polytropic and logotropic Kramers equation  if $x$ is replaced by
$v$ \cite{gen}.

The stationary solution of Eq.~(\ref{qp1}) can be written
\begin{equation}
\label{qp2} \rho=Q\left\lbrack 1-\frac{1}{2} b (\gamma-1)  \left
(x+\frac{\lambda}{\beta}\right )^{2}\right
\rbrack^{\frac{1}{\gamma-1}},
\end{equation}
where $b=\beta/Q^{\gamma-1}$ and $Q$ is determined by the
normalization condition (see below). For $\gamma>1$, the density goes
to zero at finite values of $x=x_{1},x_{2}$ (say) and it is implicitly
understood that $\rho=0$ for $x<x_{1}$ and $x>x_{2}$. For $\gamma<1$,
the density distribution is defined for all $x$ and it decreases algebraically as $x^{-2/(1-\gamma)}$. Following the
original idea of Plastino \& Plastino \cite{pp}, we look for
time-dependant solutions of the polytropic Smoluchowski equation in
the form
\begin{equation}
\label{qp3} \rho(x,t)=Q(t)\left\lbrack 1-\frac{1}{2}b(t) (\gamma-1)
(x-m(t))^{2}\right\rbrack^{\frac{1}{\gamma-1}},
\end{equation}
where $Q(t)$, $b(t)$ and $m(t)$ are functions of time. The
normalization condition $\int_{-\infty}^{+\infty}  \rho dx=1$
requires that $\gamma>-1$ and leads to the relation
\begin{equation}
\label{qp4}
b(t)=2Q(t)^{2}\mu^{2},
\end{equation}
where we have defined
\begin{equation}
\label{qp5} \mu\equiv \int_{-\infty}^{+\infty}\left\lbrack
1-(\gamma-1)y^{2} \right\rbrack^{\frac{1}{\gamma-1}}\,dy.
\end{equation}
Substituting the Ansatz (\ref{qp3}) in Eq.~(\ref{qp1}), we find
after straightforward calculations that
\begin{equation}
\label{qp6}
\dot m=-\beta m-\lambda,
\end{equation}
\begin{equation}
\label{qp7}
\dot Q=-2\mu^{2}Q^{\gamma+2}+\beta Q.
\end{equation}
Starting from a Dirac distribution $\rho(x,0)=\delta(x-x_0)$, these
equations are readily integrated and we obtain
\begin{equation}
\label{qp8} Q(t)=\left\lbrack \frac{\beta}{2\mu^{2}}
\frac{1}{1-e^{-\beta(\gamma+1)t}}\right\rbrack^{\frac{1}{\gamma+1}},
\end{equation}
\begin{equation}
\label{qp9}
m(t)=x_{0}e^{-\beta t}+\frac{\lambda}{\beta}(e^{-\beta t}-1),
\end{equation}
\begin{equation}
\label{qp10}
b(t)=2\mu^{2}Q(t)^{2}.
\end{equation}
For $\gamma\neq 0$, this returns the results obtained in \cite{pp,tb}.
However, we can now pass to the logotropic limit $\gamma\rightarrow
0$ and get
\begin{equation}
\label{qp11}
\rho(x,t)=\frac{Q(t)}{1+\frac{1}{2}b(t)(x-m(t))^{2}},
\end{equation}
\begin{equation}
\label{qp12}
Q(t)=\frac{\beta}{2\pi^{2}}\frac{1}{1-e^{-\beta t}},
\end{equation}
\begin{equation}
\label{qp13}
m(t)=x_{0}e^{-\beta t}+\frac{\lambda}{\beta}(e^{-\beta t}-1),
\end{equation}
\begin{equation}
\label{qp14}
b(t)=2\pi^{2}Q(t)^{2}.
\end{equation}
In particular for $\beta=0$, we obtain
\begin{equation}
\label{qp15} Q(t)=\frac{1}{2\pi^{2}t}, \quad m(t)=x_{0}-\lambda t,
\quad b(t)=\frac{1}{2\pi^{2}t^{2}}.
\end{equation}
Hence
\begin{equation}
\label{qp16}
\rho(x,t)=\frac{1}{2\pi^{2}t}\frac{1}{1+\frac{(x-x_{0})^{2}}{4\pi^{2}t^{2}}}.
\end{equation}

\section{The logotropic Smoluchowski-Poisson system} \label{sec_lsp}

We shall consider here the coupling between the logotropic
Smoluchowski equation and gravity. Thus, we consider the logotropic Smoluchowski-Poisson (LSP) system
\begin{equation}
\label{lsp1} \frac{\partial\rho}{\partial t}=\nabla\cdot  \left
\lbrack\frac{1}{\xi}\left (A\nabla \ln\rho+\rho\nabla\Phi \right
)\right \rbrack,
\end{equation}
\begin{equation}
\label{lsp2}
\Delta\Phi=S_{d}G\rho,
\end{equation}
in $d$-dimensions \footnote{It should be made clear that, in the
present work, we consider a self-gravitating Langevin gas with a {\it
prescribed} polytropic or logotropic equation of state. We furthermore
consider a strong friction limit in which the dynamics is governed by
the (generalized) Smoluchowski equation. This context is very
different from the works of Taruya \& Sakagami \cite{ts,tsmnras} who
consider the Hamiltonian $N$-stars system described by the
orbit-averaged-Fokker-Planck equation and show, by means of direct
numerical simulations, that the transient phases of the dynamics
(quasi-equilibrium states) can be fitted by a sequence of polytropic
distributions (Tsallis) with a time dependant index $n(t)$ or $q(t)$.
}. This can be viewed as a particular case of the polytropic
Smoluchowski-Poisson system studied in \cite{lang}, corresponding to
the index $\gamma=0$ ($n=-1$). This case must be treated separately
(i) because the study of \cite{lang} is restricted to $n\ge 0$ and
(ii) because the index $\gamma=0$ needs a special discussion as we
have seen previously. As indicated in Sec. \ref{sec_gs}, the steady
solutions of the LSP system are the same as the steady solutions of
the Euler-Poisson system with an equation of state $p=A\ln\rho$
describing logotropic stars and logotropic clusters in astrophysics
\cite{log}. Therefore, the results of
Secs. \ref{sec_lane}-\ref{sec_stab} for the steady states also apply
to logotropic stars and logotropic clusters, while the results of Sec.
\ref{sec_self} for the dynamics only apply to the LSP system.

\subsection{The logotropic Lane-Emden equation} \label{sec_lane}

For a spherically symmetric distribution, the equation of
hydrostatic equilibrium (\ref{gs4}) combined with the Gauss theorem
$d\Phi/dr=GM(r)/r^{d-1}$ can be written
\begin{equation}
\label{lane1} \frac{dp}{dr}=-\rho\frac{GM(r)}{r^{d-1}},
\end{equation}
where $M(r)=\int \rho(r',t)S_{d}r'^{d-1}dr'$ is the mass within the
sphere of radius $r$. The foregoing equation can be put in the form
\begin{equation}
\label{lane2} \frac{1}{r^{d-1}}\frac{d}{dr}\left
(\frac{r^{d-1}}{\rho} \frac{dp}{dr}\right )=-S_{d}G\rho.
\end{equation}
For the logotropic equation of state $p=A\ln\rho$, we define
\begin{equation}
\label{lane3} \rho=\rho_{0}/\theta, \qquad \xi= \left
(\frac{S_{d}G\rho_{0}^{2}}{A}\right)^{1/2}r,
\end{equation}
where $\rho_{0}$ is the central density and $\xi$ a scaled
distance.  Substituting these relations in Eq.~(\ref{lane2}), we
obtain what might be called  the logotropic Lane-Emden equation
\begin{equation}
\label{lane4} \frac{1}{\xi^{d-1}}\frac{d}{d\xi} \left
({\xi^{d-1}}\frac{d\theta}{d\xi}\right )=\frac{1}{\theta},
\end{equation}
with $\theta(0)=1$ and $\theta'(0)=0$.

\subsection{The singular logotropic sphere} \label{sec_sing}

We note that for $d>1$, the function
\begin{equation}
\label{sing1} \theta_{s}=\frac{\xi}{\sqrt{d-1}},
\end{equation}
is solution of the logotropic Lane-Emden equation. This is, however,
not a regular solution since it does not satisfy the boundary
conditions at $\xi=0$. In terms of the density, this corresponds to
a singular sphere
\begin{equation}
\label{sing2}
\rho_{s}=\left\lbrack \frac{A}{S_{d}(d-1)G}\right \rbrack^{1/2}\frac{1}{r},
\end{equation}
whose density diverges at the origin. It is interesting to compare
this {\it singular logotropic sphere} whose density profile decreases
as $r^{-1}$ for $d>1$ to the {\it singular isothermal sphere}
$\rho_{s}=\lbrack 2(d-2)/(S_{d}G\beta)\rbrack r^{-2}$ whose density
profile decreases as $r^{-2}$ for $d>2$ \cite{sc}. We note that the
total mass of the system diverges for $r\rightarrow +\infty$.
However, if we consider box-confined configurations (in a sphere of
radius $R$), the density distribution of the singular logotropic
sphere can be written
\begin{equation}
\label{sing3}
\rho_{s}=\left\lbrack \frac{M(d-1)}{S_{d}R^{d-1}}\right \rbrack \frac{1}{r}.
\end{equation}
It should be finally recalled that such singular solutions are not
limited to isothermal and logotropic distributions: polytropic spheres
with index $n>n_{3}=d/(d-2)$ (for $d>2$) and $n<-1$ (for $d>1$) also
admit singular solutions scaling like $r^{-\alpha}$ with
$\alpha=2n/(n-1)$ \cite{lang}.

\subsection{Logotropic profile in $d=1$} \label{sec_un}

\begin{figure}
\vskip1cm
\centerline{
\psfig{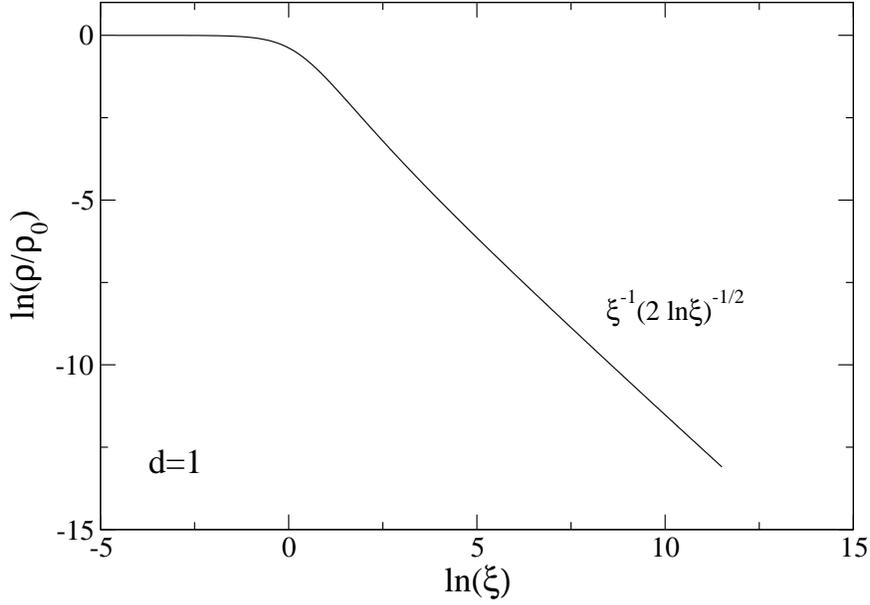}}
\caption{Density profile of a logotrope in $d=1$.}
\label{rhoxid1}
\end{figure}

In $d=1$, the logotropic Lane-Emden equation becomes
\begin{equation}
\label{un1}
\frac{d^{2}\theta}{d\xi^{2}}=\frac{1}{\theta},
\end{equation}
with $\theta(0)=1$ and $\theta'(0)=0$. This is similar to the equation
of motion of a particle in a potential $V(\theta)=-\ln\theta$ where
$\theta$ plays the role of position and $\xi$ the role of time. Using
the boundary conditions at $\xi=0$, the first integral is
\begin{equation}
\label{un2}
\frac{1}{2}\left (\frac{d\theta}{d\xi}\right )^{2}-\ln\theta=0,
\end{equation}
leading to
\begin{equation}
\label{un3}
\int_{1}^{\theta}\frac{d\phi}{\sqrt{\ln\phi}}=\sqrt{2}\xi.
\end{equation}
Performing the change of variable $\ln\phi=y^2$, we obtain the equivalent expression
\begin{equation}
\label{un4}
\sqrt{2}\int_{0}^{\sqrt{\ln\theta}}e^{y^{2}}dy=\xi.
\end{equation}
For $\xi\rightarrow +\infty$, we get
\begin{equation}
\label{un5}
\theta\sim \xi\sqrt{2\ln\xi}.
\end{equation}
Therefore, the density profile decreases like $1/(r\sqrt{\ln r})$.
The mass within a domain of size $r$ behaves as $M(r)\sim (\ln
r)^{1/2}$ for $r\rightarrow +\infty$ so that the total mass in
infinite. The density profile of a one-dimensional logotrope is
represented in Fig.~\ref{rhoxid1}.

\subsection{Asymptotic behaviors} \label{sec_as}

\begin{figure}
\vskip1cm \centerline{
\psfig{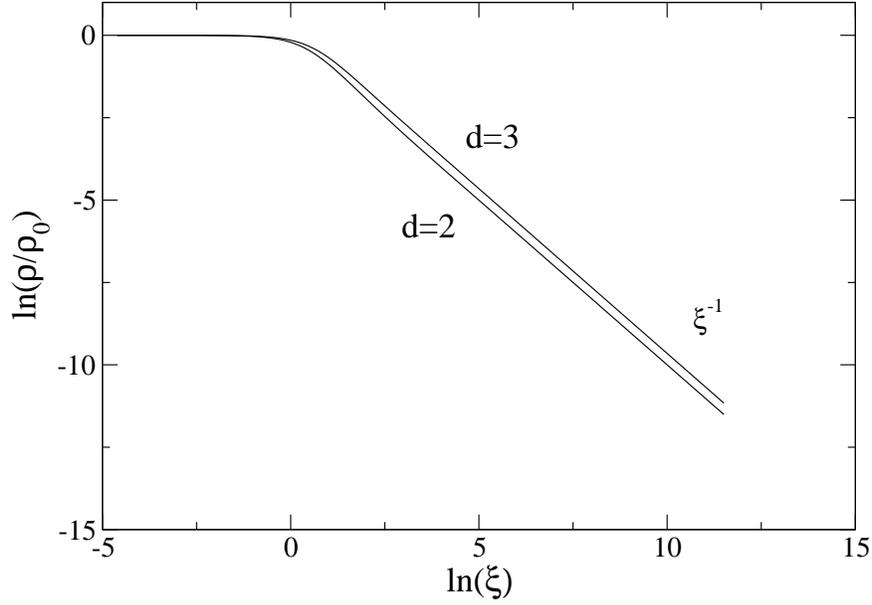}} \caption{Density
profile of a logotrope in $d=2$ and $d=3$  (the oscillations are
hardly visible but they exist).} \label{rhoxi23}
\end{figure}

We shall determine the asymptotic behaviors of the solutions of the
logotropic Lane-Emden equation (\ref{lane4}). For $\xi\rightarrow 0$, using a
Taylor expansion, we obtain
\begin{equation}
\label{as1}
\theta=1+\frac{1}{2d}\xi^{2}-\frac{1}{8d(d+2)}\xi^{4}+...\qquad (\xi\rightarrow 0).
\end{equation}
To investigate the limit $\xi\rightarrow +\infty$,  we first
perform the change of variables $t=\ln\xi$ and $\theta=\xi z$.
Equation (\ref{lane4}) then becomes
\begin{equation}
\label{as2}
\frac{d^{2}z}{dt^{2}}+d\frac{dz}{dt}=\frac{1}{z}-(d-1)z.
\end{equation}
This is similar to the damped  motion of a particle in a potential
$V(z)=-\ln z+(d-1)z^{2}/2$. For $d>1$, the potential  has a minimum
at $z_{c}=1/\sqrt{d-1}$. Thus, for $t\rightarrow +\infty$ the
``particle'' will reach this minimum, i.e. $z\rightarrow z_{c}$.
Returning to original variables, this implies that
$\theta\rightarrow \theta_{s}=\xi/\sqrt{d-1}$ for $\xi\rightarrow
+\infty$. To get the next order correction, we set $z=z_{c}+z'$
where $z'\ll z_{c}$. Keeping only terms that are linear in $z'$, we
obtain
\begin{equation}
\label{as3} \frac{d^{2}z'}{dt^{2}}+d\frac{dz'}{dt}+\left\lbrack
\frac{d^{2}-2d+2}{d-1}\right\rbrack^{2}z'=0.
\end{equation}
Noting that the discriminant $\Delta(d)=d^2-4(d^2-2d+2)^2/(d-1)^2$
of the associated quadratic equation is always negative for $d>1$,
we find that
\begin{equation}
\label{as4}
z'=e^{-dt/{2}}\cos\left (\frac{\sqrt{-\Delta}}{2}t+\delta\right).
\end{equation}
Returning to original variables, we obtain the asymptotic behavior
\begin{equation}
\label{as5} \theta\sim \theta_{s}(\xi)\left\lbrace
1+\frac{C}{\xi^{d/2}}\cos \left
(\frac{\sqrt{-\Delta}}{2}\ln\xi+\delta\right )\right\rbrace,
\qquad (\xi\rightarrow +\infty).
\end{equation}
As in the case of isothermal and polytropic distributions
\cite{sc,lang}, the density profile of a logotrope presents damped
oscillations around the singular solution. The density profiles of
a logotrope in $d=2$ and $d=3$ are represented in
Fig.~\ref{rhoxi23}.  For these dimensions $\Delta(2)=-12$ and
$\Delta(3)=-16$.

\subsection{The Milne variables} \label{sec_milne}

\begin{figure}
\vskip1cm \centerline{
\psfig{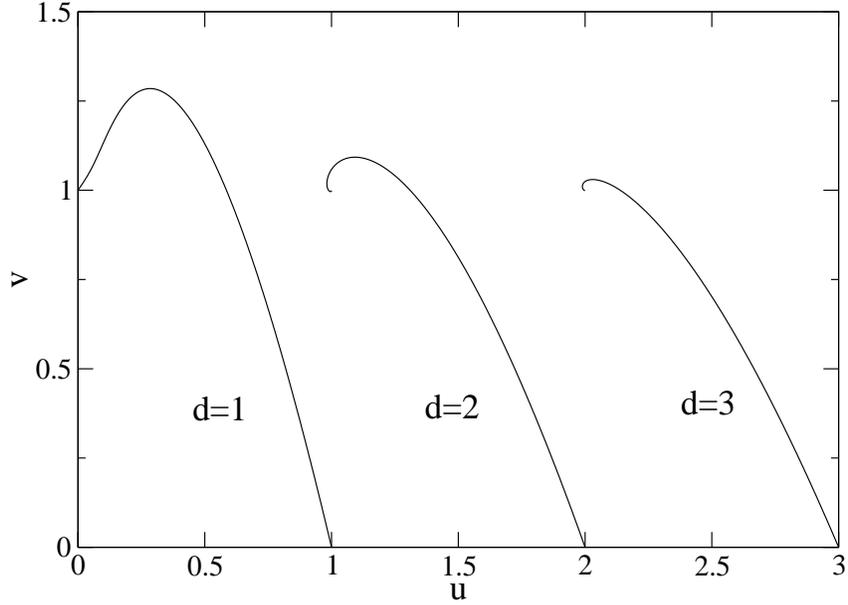}} \caption{The
solutions of the logotropic Lane-Emden equation  in the Milne
plane $(u,v)$ for $d=1$, $d=2$ and $d=3$.} \label{milne123}
\end{figure}

As in the case of polytropes \cite{lang}, it is convenient to introduce the variables
\begin{equation}
\label{milne1}
u=\frac{\xi}{\theta \theta'}, \qquad v=\frac{\xi\theta'}{\theta},
\end{equation}
which are the appropriate forms of the Milne variables \cite{chandra}
in the present context. Using the logotropic Lane-Emden equation (\ref{lane4})
we easily derive
\begin{equation}
\label{milne2}
\frac{1}{u}\frac{du}{d\xi}=\frac{1}{\xi}(d-v-u),
\end{equation}
\begin{equation}
\label{milne3}
\frac{1}{v}\frac{dv}{d\xi}=\frac{1}{\xi}(2-d+u-v),
\end{equation}
so that the function $v(u)$  satisfies the first order
differential equation
\begin{equation}
\label{milne4}
\frac{u}{v}\frac{dv}{du}=-\frac{u-v-d+2}{u+v-d}.
\end{equation}
As for isothermal and polytropic spheres, the reduction of a second
order differential equation (\ref{lane4}) to a first order
differential equation (\ref{milne4}) is a consequence of the homology
theorem \cite{chandra}. The solution curve in the Milne plane is
parameterized by $\xi$ going from $0$ to $+\infty$. It is represented
in Fig.~\ref{milne123} and forms a spiral for $d>1$. The points of
horizontal tangent correspond to $2-d+u-v=0$ and the points of
vertical tangent to $d-v-u=0$. They coincide for $u_{s}=d-1$ and
$v_{s}=1$ which corresponds to the singular logotropic sphere
(\ref{sing1}). On the other hand, the curve starts at $(d,0)$ for
$\xi=0$ with the slope $(dv/du)_{0}=-(d+2)/d$ and rolls up to the
singular logotropic sphere (when $d>1$) for $\xi\rightarrow
+\infty$. For $d=1$, the solution curve tends to $(0,1)$ for
$\xi\rightarrow +\infty$. It does not make a spiral but it presents a
maximum at some point.

\subsection{The series of equilibria} \label{sec_series}

\begin{figure}
\vskip1cm
\centerline{
\psfig{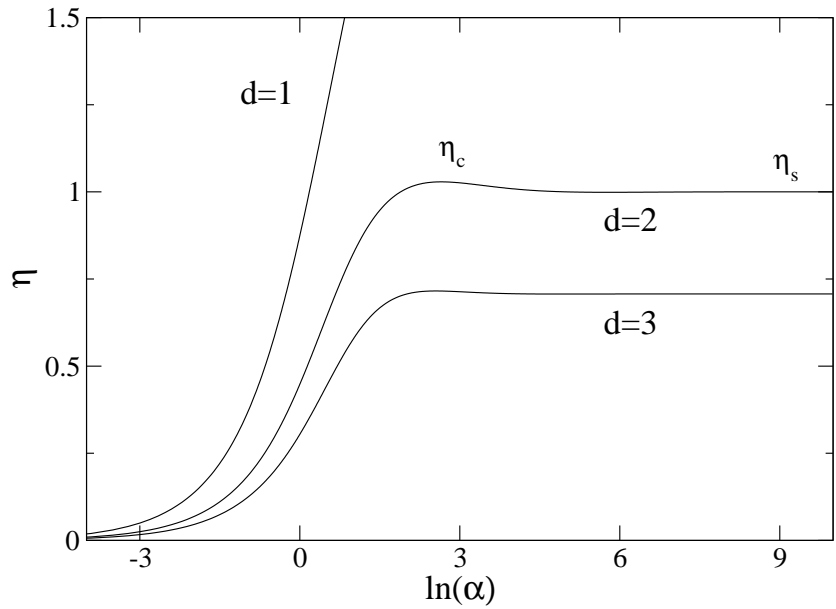}}
\caption{Series of equilibria for box-confined logotropes in $d=1$, $d=2$ and $d=3$.}
\label{alphaeta123}
\end{figure}

Since the density profile of logotropic distributions in $d>1$
decreases as $r^{-1}$ at large distances, the total mass is
infinite (the total mass is also infinite in $d=1$). Therefore, we
shall enclose these distributions in a spherical box of radius $R$
as in the case of isothermal and polytropic spheres
\cite{sc,lang}. For box-confined logotropes, the solution of
Eq.~(\ref{lane4}) is terminated by the box at a normalized radius
\begin{equation}
\label{series1}
\alpha=\left (\frac{S_{d}G\rho_{0}^{2}}{A}\right )^{1/2}R.
\end{equation}
The total mass of the configuration can be written
\begin{eqnarray}
\label{series2} M=\int_{0}^{R}\rho S_{d}r^{d-1}\,dr=S_{d}\rho_{0}
\left (\frac{A}{S_{d}G\rho_{0}^{2}}\right )^{d/2}
\int_{0}^{\alpha}\frac{1}{\theta}\xi^{d-1}\,d\xi\nonumber\\
=S_{d}\rho_{0}\left (\frac{A}{S_{d}G\rho_{0}^{2}}\right )^{d/2}
\int_{0}^{\alpha}  \frac{d}{d\xi}\left ({\xi^{d-1}}
\frac{d\theta}{d\xi}\right )\,d\xi=\left (\frac{S_{d} A}{G}\right
)^{1/2} R^{d-1}\theta'(\alpha).
\end{eqnarray}
It is  therefore natural to introduce the dimensionless parameter
\begin{eqnarray}
\label{series3}
\eta=M\left (\frac{G}{S_{d}A}\right )^{1/2}\frac{1}{R^{d-1}}.
\end{eqnarray}
Then, the series of equilibria is defined by
\begin{eqnarray}
\label{series4}
\eta=\theta'(\alpha).
\end{eqnarray}
This relation can be interpreted in different manners. For
example, if we fix $A$ and $R$, the parameter $\alpha$ is
proportional to the central density $\rho_{0}$ and the parameter
$\eta$ to the mass $M$. Therefore, Eq.~(\ref{series4}) can be
interpreted as the mass-central density relation. Alternatively,
if we fix $M$ and $R$, then the parameter $\eta$ is related to
$A=T_{eff}$ which is interpreted as a ``generalized temperature''.
Thus Eq.~(\ref{series4}) gives the relation between the central
density and the generalized temperature. In terms of the Milne
variables, the relation (\ref{series4})  can be rewritten
\begin{eqnarray}
\label{series5}
\eta=\left ( \frac{v_{0}}{u_{0}}\right )^{1/2},
\end{eqnarray}
where, by definition, $u_{0}=u(\alpha)$ and $v_{0}=v(\alpha)$ are
the values of the Milne variables at the box radius. The series of
equilibria (\ref{series4}) is plotted in Fig.~\ref{alphaeta123}
and it presents damped oscillations for $d>1$ (for $d=1$ the curve
is monotonic). Similar oscillations are encountered for isothermal
spheres in Newtonian gravity \cite{aa1} and general relativity
\cite{aa2}. For $\alpha\rightarrow +\infty$, we recover the
singular sphere with $\eta_{s}=1/\sqrt{d-1}$. The control
parameter $\eta(\alpha)$ is extremum for $\alpha_{c}$ such that
$d\eta/d\alpha=0$. Using Eqs. (\ref{milne2})-(\ref{milne3}), we
find that this condition is equivalent to
\begin{eqnarray}
\label{series6}
u_{0}=d-1=u_{s}.
\end{eqnarray}
The number of extrema can be obtained by a simple graphical
construction in the Milne plane. They are determined by the intersection between the $(u,v)$ curve and the line $u=u_{s}$. Since the straight line $u=u_{s}$
passes by the center of the spiral, there is an infinite number of
extrema.  The critical value of $\eta$ corresponding to the first
maximum is $\eta_{c}=0.715657...$ in $d=3$ and $\eta_{c}=1.028728...$ in $d=2$.
Clearly, there is no steady state for $\eta>\eta_{c}$. Similar results
are found for isothermal systems \cite{aa1,sc}. Note, however, that
the series of equilibria $\eta(\alpha)$ for logotropes already
presents oscillations in $d=2$ unlike isothermal spheres for which
oscillations appear for $d>2$.

\subsection{Stability analysis} \label{sec_stab}

We shall now study the dynamical stability of logotropic spheres
by determining whether they are {\it minima} of the generalized
free energy functional (\ref{logs3}) at fixed mass. As shown in
\cite{gen}, a stationary solution of the generalized
Smoluchowski-Poisson system (\ref{gs1})-(\ref{gs2}) is linearly
dynamically stable if, and only if, it is a minimum of the functional
$F$ defined by Eq.~(\ref{gs3}). In that Brownian context, $F$ is
interpreted as a Lyapunov functional or as a (generalized) free
energy, and its minimization corresponds to an effective
thermodynamical stability criterion in the canonical ensemble. On the
other hand, a stationary solution of the barotropic Euler-Poisson
system (\ref{gs5})-(\ref{gs7}) is nonlinearly dynamically stable if it
is a minimum of $F$. In that Euler context, $F$ is related to the
energy functional of the barotropic gas ${\cal W}$ (see
\cite{lang,intT,antofirst} for details).  Therefore, our stability
analysis has applications for these two different systems: a
Brownian gas described by nonlinear Fokker-Planck equations and a
``normal'' gas described by the Euler equations. Since the
following stability analysis is very similar to that developed in
\cite{lang} for polytropic distributions, we shall only give the
main lines of the analysis and refer to \cite{lang} for more
details.

\begin{figure}
\vskip1cm \centerline{
\psfig{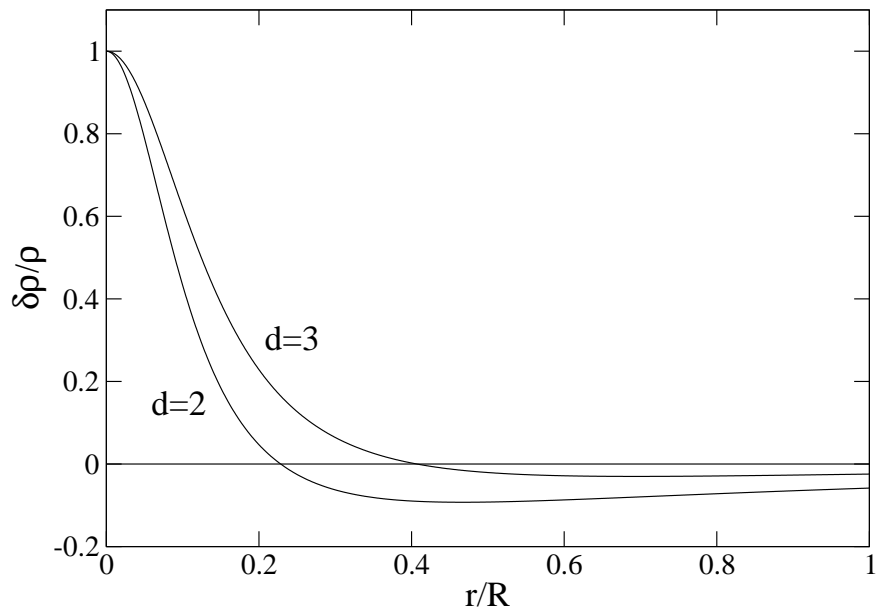}}
\caption{Perturbation profile corresponding to the first  mode of
instability at $\eta_{c}$ in  $d=2$ and $d=3$.} \label{deltarho23}
\end{figure}

The second order variations of $F$  are given by
\begin{eqnarray}
\label{stab1} \delta^{2}F=\frac{A}{2}\int
\frac{(\delta\rho)^{2}}{\rho^{2}}\, d{\bf r}+ \frac{1}{2}\int
\delta\rho \delta\Phi \, d{\bf r}.
\end{eqnarray}
Restricting ourselves to spherically symmetric perturbations,
introducing the quantity $q(r)$ through the defining relation
\begin{eqnarray}
\label{stab2}
\delta\rho=\frac{1}{S_{d}r^{d-1}}\frac{dq}{dr},
\end{eqnarray}
and using integrations by parts, we can put the second order
variations of $F$  in the quadratic form
\begin{eqnarray}
\label{stab3} \delta^{2}F=-\frac{1}{2}\int_{0}^{R}q\frac{d}{dr} \left
(\frac{A}{S_{d}\rho^{2}r^{d-1}}\frac{d}{dr}+\frac{G}{r^{d-1}}\right
)q \, dr.
\end{eqnarray}
We are thus led to considering the eigenvalue equation
\begin{eqnarray}
\label{stab4} \frac{d}{dr}\left
(\frac{A}{S_{d}\rho^{2}r^{d-1}}\frac{d}{dr}
+\frac{G}{r^{d-1}}\right )q_{\lambda}(r)=\lambda q_{\lambda}(r),
\end{eqnarray}
and determine when, in the series of equilibria, the eigenvalues
pass from positive (stable) to negative (unstable) values. To that
purpose, it is sufficient to determine the point of marginal
stability corresponding to $\lambda=0$. Introducing the dimensionless quantities of Sec. \ref{sec_lane}, we have to solve
\begin{eqnarray}
\label{stab5} {\cal L}F\equiv \frac{d}{d\xi}\left
(\frac{\theta^{2}}{\xi^{d-1}} \frac{dF}{d\xi}\right
)+\frac{F}{\xi^{d-1}}=0,
\end{eqnarray}
with the boundary conditions $F(0)=0$ and $F(\alpha)=0$ implied by the conservation of mass. Noting that
\begin{eqnarray}
\label{stab6}
{\cal L}(\xi^{d-1}\theta')=2\theta', \qquad {\cal L}\left (\frac{\xi^{d}}{\theta}\right )=2(d-1)\theta',
\end{eqnarray}
we find that the solution of Eq.~(\ref{stab5}) satisfying $F(0)=0$
is
\begin{eqnarray}
\label{stab7}
F=a\left\lbrack \frac{\xi^{d}}{\theta}-(d-1)\xi^{d-1}\theta'\right\rbrack,
\end{eqnarray}
where $a$ is an arbitrary constant (due to the linearity of the
eigenvalue equation).  The critical values of $\alpha$ where an
eigenvalue becomes zero are determined by the condition
$F(\alpha)=0$. Using Eq.~(\ref{stab7}) and introducing the Milne
variables (\ref{milne1}), we find that this is equivalent to
$u_{0}=d-1=u_{s}$. Comparing with Eq.~(\ref{series6}), we find that a
new eigenvalue becomes zero each time that $\eta$ is extremum. As
a result, the series of equilibria becomes unstable at the point
of maximum $\eta$, previously denoted $\eta_c$. This result is
consistent with the turning point argument of Poincar\'e (see
\cite{lang} for details).  The perturbation triggering the
instability at the marginal point is given by
\begin{eqnarray}
\label{stab8}
\frac{\delta\rho}{\rho_{0}}=\frac{1}{S_{d}\xi^{d-1}}\frac{dF}{d\xi}.
\end{eqnarray}
In terms of the Milne variables, this can be rewritten
\begin{eqnarray}
\label{stab9}
\frac{\delta\rho}{\rho}=\frac{a}{S_{d}}(v_{s}-v).
\end{eqnarray}
Hence, the number of oscillations of the perturbation profile
$\delta\rho(\xi)$ can be easily  determined by graphical
constructions. The values of $\xi$ where $\delta\rho(\xi)$
vanishes are determined by the intersection between the $(u,v)$
curve and the line $v=v_{s}$. Of course, we have to consider only
values of $\xi\le \alpha$. At the point of instability $\eta_{c}$,
the perturbation has only one node (see Fig.~\ref{deltarho23}).

We can also study the linear dynamical stability of a stationary
solution of the LSP system and show the equivalence with the
stability criterion based on the minimization of the free energy.
Linearizing the generalized SP system around a stationary solution
and writing the perturbation in the form $\delta\rho\sim
e^{\lambda t}$ we finally arrive at the eigenvalue equation
\cite{gen}:
\begin{eqnarray}
\label{stab10} \frac{d}{dr}\left (\frac{p'(\rho)}{S_{d}\rho
r^{d-1}} \frac{dq}{dr}\right
)+\frac{Gq}{r^{d-1}}=\frac{\lambda\xi}{S_{d}\rho r^{d-1}}q.
\end{eqnarray}
Specializing to the case of a logotropic equation of state $p=A\ln\rho$, we obtain
\begin{eqnarray}
\label{stab11} \frac{d}{dr}\left (\frac{A}{S_{d}\rho^{2} r^{d-1}}
\frac{dq}{dr}\right
)+\frac{Gq}{r^{d-1}}=\frac{\lambda\xi}{S_{d}\rho r^{d-1}}q.
\end{eqnarray}
This equation is similar to Eq.~(\ref{stab4}) and they coincide at the
point of marginal stability. Therefore, the stability threshold is the
same in the two approaches and the above results concerning the form
of the perturbation apply. Finally, if we consider the linear
dynamical stability of a steady solution of the logotropic
Euler-Poisson system (\ref{gs5})-(\ref{gs7}), we obtain an eigenvalue
equation of the form (\ref{stab11}) where $\lambda\xi$ is replaced by
$\lambda^{2}$ \cite{virial}. Once again, the point of marginal
dynamical stability coincides with the point where the energy
functional ceases to be a minimum and becomes a saddle
point. Therefore, the conditions of linear and nonlinear dynamical
stability for the Euler-Poisson system coincide.

\subsection{Self-similar solutions of the polytropic and logotropic
Smo\-lu\-chow\-ski-Poisson system} \label{sec_self}

When no stable equilibrium state exists, the polytropic
Smoluchowski-Poisson system admits self-similar solutions describing a
gravitational collapse. These solutions have been given in \cite{lang}
for a polytropic index $n\ge 0$. As we shall see, the situation is
different, and richer, when $n\le 0$ (this case includes in particular
the logotropes $n=-1$). We restrict our analysis to a space of
dimension $d>2$ (the dimension $d=2$ is critical and requires a
particular and intricate treatment; see \cite{sc} for isothermal
systems).

If we introduce the function $s(r,t)=M(r,t)/r^{d}$, which scales like
the density $\rho(r,t)$, we can rewrite the polytropic
Smoluchowski-Poisson system (\ref{polys1})-(\ref{lsp2}) as a single
differential equation \cite{lang}:
\begin{eqnarray}
\label{self1} \frac{\partial s}{\partial t}=\Theta \left (
r\frac{\partial s}{\partial r}+ds\right )^{1/n}\left
(\frac{\partial^{2}s}{\partial r^{2}}+\frac{d+1}{r}\frac{\partial
s}{\partial r}\right )+\left (r\frac{\partial s}{\partial
r}+ds\right )s,
\end{eqnarray}
where we have defined $A=K\gamma$ and introduced an effective
temperature $\Theta=A/S_{d}^{1/n}$. We note that this equation passes
to the limit for $n=-1$.  We look for self-similar solutions of the
form
\begin{eqnarray}
\label{self2}
s(r,t)=\rho_{0}(t) S\left( \frac{r}{r_{0}(t)}\right ),
\end{eqnarray}
where $\rho_{0}(t)$ is proportional to the central density.  The
scaling solution (\ref{self2}) only holds in the central region of
the cluster. If we assume that all the terms scale the same in
Eq.~(\ref{self1}), which needs not be the case (see below), we
find that
 \begin{eqnarray}
\label{self3}
\Theta \frac{\rho_{0}^{1/n+1}}{r_{0}^{2}}\sim \rho_{0}^{2}.
\end{eqnarray}
We can then define the scaling radius by
\begin{eqnarray}
\label{self4}
r_{0}=\left (\frac{\Theta}{\rho_{0}^{1-1/n}}\right )^{1/2}.
\end{eqnarray}
In that case, the relation between the typical central  density
and the typical core radius is
\begin{eqnarray}
\label{self5}
\rho_{0}\sim r_{0}^{-\alpha}, \qquad {\rm with}\quad \alpha=\frac{2n}{n-1}.
\end{eqnarray}
Substituting these relations in Eq.~(\ref{self1}), we find that
\begin{eqnarray}
\label{self6} \frac{d\rho_{0}}{dt}\left
(S+\frac{1}{\alpha}xS'\right )= \rho_{0}^{2}\left\lbrack
(xS'+dS)^{1/n}\left (S''+\frac{d+1}{x}S'\right
)+(xS'+dS)S\right\rbrack,
\end{eqnarray}
where we have set $x=r/r_{0}$. This implies that
$(1/\rho_{0}^{2})(d\rho_{0}/dt)$ is a constant that we arbitrarily
set equal to $\alpha$. This leads to
\begin{eqnarray}
\label{self7}
\rho_{0}(t)=\frac{1}{\alpha}(t_{coll}-t)^{-1},
\end{eqnarray}
so that the central density becomes infinite in a finite time
$t_{coll}$.  The scaling equation now reads
\begin{eqnarray}
\label{self8}
\alpha S+xS'=(xS'+dS)^{1/n}\left (S''+\frac{d+1}{x}S'\right )+(xS'+dS)S.
\end{eqnarray}
For $x\rightarrow +\infty$, we have the scaling $S(x)\sim
x^{-\alpha}$ so that the density behaves as $\rho\sim r^{-\alpha}$
for $r\rightarrow +\infty$. At $t=t_{coll}$ the density profile is
proportional to $1/r^{\alpha}$.

However, we could be in a situation in which the collapse  is
dominated by the gravitational drift like in the case of cold
systems where $\Theta=0$. In that case, the dynamical equation
(\ref{self1}) reduces to
\begin{eqnarray}
\label{self9}
\frac{\partial s}{\partial t}\simeq \left (r\frac{\partial s}{\partial r}+ds\right )s.
\end{eqnarray}
This equation with $\Theta=0$ has been solved in \cite{sc}.  The
collapse evolution is self-similar and the relation between the
typical central density and the typical core radius is
\begin{eqnarray}
\label{self10}
\rho_{0}\sim r_{0}^{-\alpha}, \qquad {\rm with}\quad \alpha=\frac{2d}{d+2}.
\end{eqnarray}
We expect that this ``cold regime'' will prevail over the regime
where the diffusion and the gravitational terms scale the same way
if  $(\rho_{0})_{\Theta=0}\gg (\rho_{0})_{\Theta\neq 0}$, i.e. if
\begin{eqnarray}
\label{self11}
\frac{2d}{d+2}>\frac{2n}{n-1}.
\end{eqnarray}
Indeed, one expects on the basis of free energy arguments that the
most natural evolution is the one which leads to the most
efficient  collapse of the core. This is obtained by selecting the
largest value of $\alpha$ between Eqs.~(\ref{self5}) and
(\ref{self10}).

\begin{figure}
\vskip1cm \centerline{\psfig{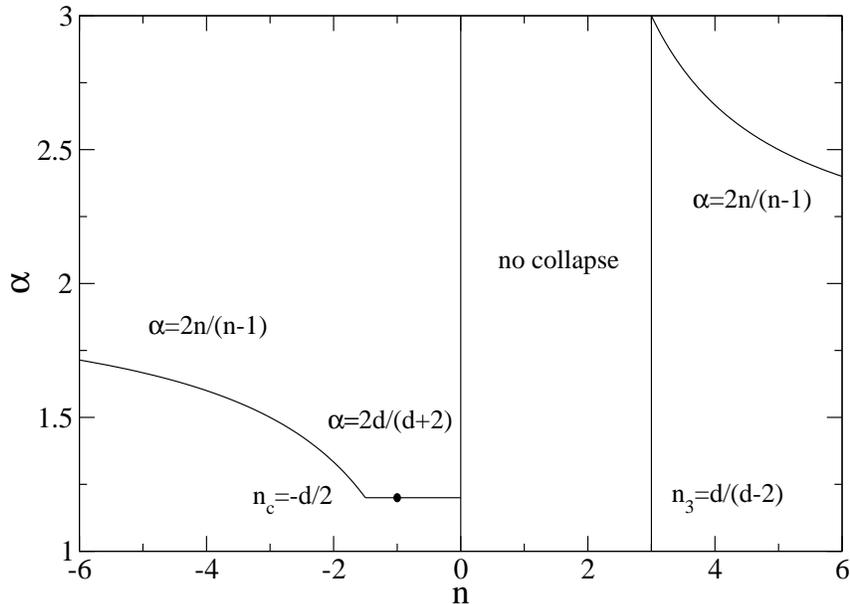}}
\caption{Scaling exponent of the self-similar collapse as  a
function of the polytropic index (the curve is represented for
$d=3$).} \label{exp}
\end{figure}

Let us consider the range of validity of this inequality:

(i) The case $n\ge 0$ has been considered in \cite{lang}.  It is
shown that self-similar solutions exist only for $n>n_{3}=d/(d-2)$
(otherwise the system tends to a complete polytrope). We note that
inequality (\ref{self11}) is never satisfied in that case so that
the scaling exponent is $\alpha=2n/(n-1)$ independant on the
dimension. The results of \cite{lang} are therefore unaltered.

(ii) For $n<0$, the inequality (\ref{self11}) is equivalent to
$n_{c}=-d/2\le n< 0$ where $n_{c}=-d/2$ is a new critical index
which does not seem to have been introduced before. In that case, the
collapse of the polytropic SP system is similar to the case $\Theta=0$
(cold system) and the scaling exponent is $\alpha=2d/(d+2)$. The
corresponding scaling profile is known analytically in an implicit
form \cite{sc}. Finally, for $n<n_{c}=-d/2$, the scaling exponent is
$\alpha=2n/(n-1)$. In particular, for the logotropes with $n=-1$ in
$d>2$, the scaling exponent is $\alpha=2d/(d+2)$.  Therefore, the
density profile of a collapsed logotrope scales as $\rho\sim
r^{-2d/(d+2)}$ while the density profile of an equilibrium logotrope
scales as $\rho\sim r^{-1}$ for $r\rightarrow +\infty$ (see
Sec. \ref{sec_as}). This situation differs from the case of isothermal
and polytropic systems with $n\ge 0$ where the collapse exponent
$\alpha=2n/(n-1)$ is the same as the exponent controlling the decay of
the equilibrium density profile at large distances \cite{sc,lang}.

\begin{figure}
\vskip1cm \centerline{
\psfig{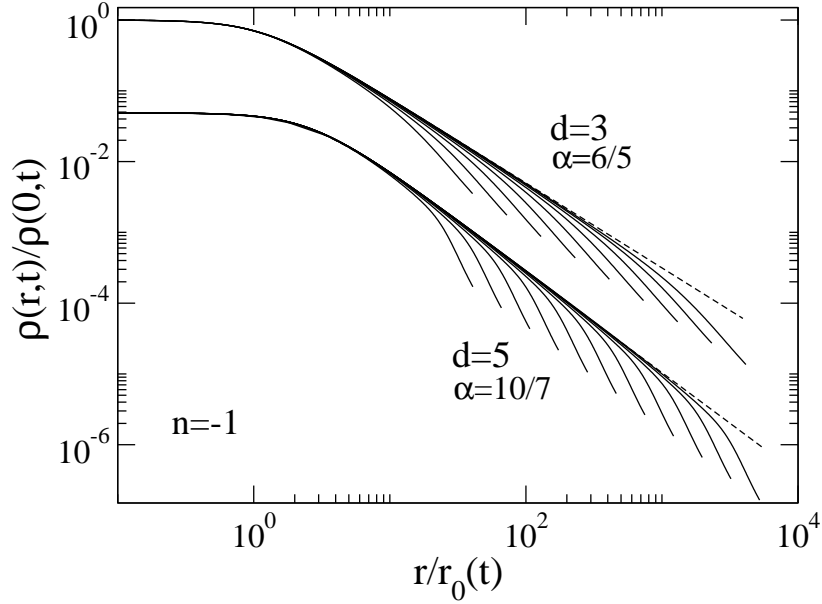}}
\caption{Data collapse for $n=-1$ obtained by solving the
dynamical equation (\ref{self1}). Upper curve: $d=3$,
$\alpha=2d/(d+2)=6/5$. Lower curve: $d=5$, $\alpha=2d/(d+2)=10/7$;
the curve has been shifted by an arbitrary factor. The invariant
profile (dashed line) corresponds to $\Theta=0$. It is known
analytically in implicit form \cite{sc}.} \label{cle1}
\end{figure}

\begin{figure}
\vskip1.3cm \centerline{
\psfig{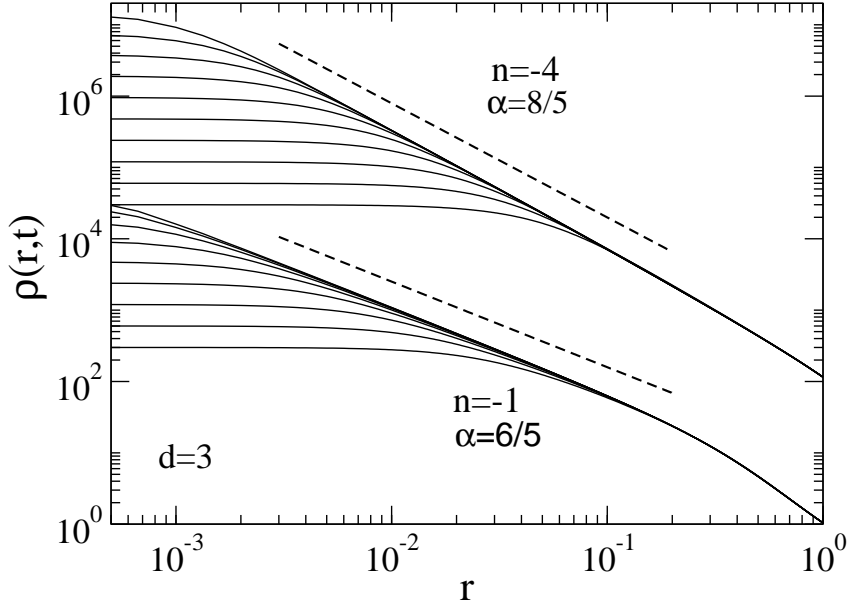}} \caption{The
density profile (divided by $S_{d}$)  is plotted for successive
times so that the central density increases by a factor $2$ at
each time, starting from a central density $300$. The dimension is
$d=3$.  Upper curve: $n=-4<n_{c}=-d/2$ so that
$\alpha=2n/(n-1)=8/5$ (regular scaling); the figure has been
shifted by a factor $100$ for clarity. Lower curve:  $n=-1>n_{c}$
so that $\alpha=2d/(d+2)=6/5$ (scaling at $\Theta=0$). The dashed
lines are the theoretical slopes.} \label{cle2}
\end{figure}

The above results are summarized in Fig.~\ref{exp}. In
Fig.~\ref{cle1}, we show the data collapse obtained by solving the
dynamical equation (\ref{self1}) with $n=-1$ (logotrope) in
dimensions $d=3$ and $d=5$. The scaling exponent agrees with its
predicted theoretical value (\ref{self10}) corresponding to a
pressureless collapse at $\Theta=0$. The scaling profile is also
in good agreement with the analytical profile at $\Theta=0$
obtained in \cite{lang} in implicit form. In Fig.~\ref{cle2}, we
show the time evolution of the density profile during the collapse
obtained by solving the dynamical equation (\ref{self1}) in
dimension $d=3$ for $n=-1>n_{c}=-3/2$ and $n=-4<n_{c}=-3/2$. The
exponents are in good agreement with the theoretical predictions
(\ref{self10}) and (\ref{self5}) respectively.

\section{Conclusion} \label{sec_conclusion}

In this paper, we have studied a special class of nonlinear
mean-field Fokker-Planck equations introduced in \cite{gen},
corresponding to a logotropic equation of state. This can be
viewed as a limiting form of polytropic equation of state with
$\gamma=0$. In the language of generalized thermodynamics, this
corresponds to a Tsallis distribution with $q=0$. Special
attention has been given to the logotropic Smoluchowski-Poisson
system where we have studied the steady states, their stability
and, when unstable, the resulting gravitational collapse. In
particular, we have shown that for polytropic indices $-d/2\le
n<0$ in $d>2$ (including the logotropes $n=-1$), the dynamics
below the critical effective temperature $\Theta<\Theta_c$ {\it
coincides} with the dynamics at $\Theta=0$. This is similar to the
coarsening dynamics of spin systems strictly below the
ferromagnetic critical temperature, which is controlled by the
zero temperature fixed point \cite{bray}. This work completes
previous investigations of the authors who studied the generalized
Smoluchowski-Poisson system for an isothermal equation of state
\cite{crs,sc}, a polytropic equation of state with $n\ge 0$
\cite{lang}, a Fermi-Dirac equation of state \cite{iso} and a Bose
statistics in velocity space \cite{bose}. Our results can also have
applications for the chemotactic aggregation of bacterial populations
in biology that is governed by similar drift-diffusion equations
\cite{iso,jtype,tr}.


\begin{thebibliography}{7}
%
\addcontentsline{toc}{section}{References}


\bibitem{collpap}  {\small Special issue of Physica A {\bf  340},
Issue 1-3, edited by G. Kaniadakis and M. Lissia (2004).}

\bibitem{frankbook}  {\small  T.D. Frank, {\it Nonlinear Fokker-Planck Equations: Fundamentals and Applications}, Springer-Verlag, 2005.}

\bibitem{blks}  {\small  P.H. Chavanis, Physica A {\bf 332}, 89 (2004).}


\bibitem{tsallis}  {\small C. Tsallis, J. Stat. Phys. {\bf 52}, 479
    (1988).}

\bibitem{pp}  {\small A.R. Plastino and A. Plastino, Physica A {\bf 222}, 347
    (1995).}

\bibitem{kaniadakis}  {\small G. Kaniadakis, Physica A {\bf 296}, 405 (2001).}

\bibitem{frank}  {\small T.D. Frank, Physics Lett. A {\bf 290}, 93 (2001).}

\bibitem{gen}  {\small P.H. Chavanis, Phys. Rev. E {\bf 68}, 036108 (2003).}

\bibitem{next03}  {\small  P.H. Chavanis, Physica A {\bf 340}, 57 (2004).}

\bibitem{banach}  {\small P.H. Chavanis, Banach Center Publ. {\bf 66}, 79 (2004).}

\bibitem{lemou}  {\small  P.H. Chavanis, P. Lauren\c{c}ot, M. Lemou,
Physica A \textbf{341}, 145 (2004).}

\bibitem{log}  {\small  D.E. McLaughlin and R.E. Pudritz, Astrophys.
J.  {\bf 476}, 750 (1997).}

\bibitem{tb}  {\small  C. Tsallis and D.J. Bukman, Phys. Rev. E  {\bf 54}, R2197 (1996).}

\bibitem{ts}   {\small A. Taruya and M. Sakagami,  Physica A {\bf 318}, 387 (2003).}

\bibitem{lang}  {\small  P.H. Chavanis and C. Sire, Phys. Rev. E {\bf 69}, 016116 (2004).}

\bibitem{crs}  {\small P.H. Chavanis, C. Rosier, and C. Sire, Phys. Rev. E {\bf  66}, 036105 (2002). }

\bibitem{sc}  {\small C. Sire and P.H. Chavanis, Phys. Rev. E {\bf 66}, 046133 (2002).}

\bibitem{iso}  {\small P.-H. Chavanis, M. Ribot, C. Rosier, and C. Sire,
Banach Center Publ. {\bf 66}, 103 (2004).}

\bibitem{hmf}  {\small  P.H. Chavanis, J.
Vatteville, and F. Bouchet,  Eur. Phys. J. B {\bf  46}, 61 (2005).
}

\bibitem{multi}  {\small  J. Sopik, C. Sire, and P.H.
Chavanis,  Phys. Rev. E {\bf  72}, 026105 (2005). }

\bibitem{bose}  {\small  J. Sopik, C. Sire, and P.H.
Chavanis,  Phys. Rev. E {\bf  74}, 011112 (2006). }

\bibitem{virial}  {\small  P.H.
Chavanis and C. Sire,  Phys. Rev. E {\bf 73}, 066103 (2006); {\it ibid} {\bf 73}, 066104 (2006). }

\bibitem{tr}  {\small  P.H.
Chavanis, preprint [physics/0607020]. }


\bibitem{hb}  {\small   P.H.
Chavanis,   Physica A {\bf 361}, 55 (2006); {\it ibid.} {\bf 361}, 81 (2006).}

\bibitem{cras}  {\small   P.H.
Chavanis,   C. R. Physique {\bf 7}, 318 (2006).}

\bibitem{bt}  {\small J. Binney and S. Tremaine,
{\it Galactic Dynamics} (Princeton Series in Astrophysics, 1987).}

\bibitem{jtype}  {\small P.H. Chavanis,  Eur. Phys. J. B {\bf 52}, 433 (2006). }

\bibitem{ks}  {\small E. Keller, L.A. Segel, J. theor. Biol. {\bf 26}, 399 (1970).}

\bibitem{jaeger}  {\small W. J\"ager, S. Luckhaus, Trans. Am. Math. Soc.  {\bf 329}, 819 (1992).}

\bibitem{super}  {\small P.H. Chavanis, Physica A {\bf 359}, 177 (2006).}

\bibitem{intT}  {\small P.H. Chavanis and C. Sire, Physica A {\bf 356}, 419 (2005).}

\bibitem{tsmnras}   {\small A. Taruya and M. Sakagami,  Mon. Not. R. Astron. Soc. {\bf 364}, 990 (2005).}

\bibitem{chandra}  {\small S. Chandrasekhar,
{\it An Introduction to the Theory of Stellar Structure} (Dover, New York, 1939).}

\bibitem{aa1}  {\small P.H. Chavanis, Astron. Astrophys.  {\bf 381}, 340 (2002).}

\bibitem{aa2}  {\small P.H. Chavanis, Astron. Astrophys.  {\bf 381}, 709 (2002).}

\bibitem{antofirst}  {\small P.H. Chavanis, Astron. Astrophys. {\bf 451}, 109 (2006).}

\bibitem{bray}   {\small A. J. Bray, Adv. Phys. {\bf 43}, 357 (1994).}





\end{thebibliography}
\end{document}